\documentclass[trackchanges,
amsmath,
preprint2]{aastex701}

\newcommand{\vs}{{V_s}}
\newcommand{\vo}{{V_o}}

\begin{document}

\title{The Enhancement of Ion Heating in Kinetic, Anti-Parallel Reconnection in the Presence of a Flow Shear}

\author[0000-0002-2160-7288]{Colby C. Haggerty}
\affiliation{Institute for Astronomy, University of Hawai`i, Honolulu, HI 96822, USA}
\email{colbyh@hawaii.edu}

\author[0000-0001-5796-2807]{Derek Sikorski}
\affiliation{Institute for Astronomy, University of Hawai`i, Honolulu, HI 96822, USA}
\email{dsikors@hawaii.edu}

\author[0000-0003-1861-4767]{Michael A. Shay}
\affiliation{Bartol Research Institute, Department of Physics \& Astronomy, University of Delaware, Newark, DE, 19716, USA }
\email{shay@udel.edu}

\author[0000-0002-6924-9408]{Tai D. Phan}
\affiliation{Space Sciences Laboratory, University of California, Berkeley, CA 94720, USA}
\email{phan@ssl.berkeley.edu}

\author[0000-0002-5938-1050]{Paul A Cassak}
\affiliation{Department of Physics and Astronomy and the Center for KINETIC Plasma Physics, West Virginia University, Morgantown, West Virginia 26506, USA}
\email{pcassak@clemson.edu}

\author[0000-0002-7836-7078]{Giulia Murtas}
\affiliation{Institute for Astronomy, University of Hawai`i, Honolulu, HI 96822, USA}
\email{gmurtas@hawaii.edu}

\author[0000-0003-4832-7638]{Prayash S. Pyakurel}
\affiliation{Space Sciences Laboratory, University of California, Berkeley, CA 94720, USA}
\email{pspyakurel@berkeley.edu}

\begin{abstract}

We investigate the kinetic effects of upstream, magnetic field-aligned, flow shear on anti-parallel magnetic reconnection using 2.5D Particle-In-Cell simulations.
Our results demonstrate that flow shear significantly alters the reconnection process, leading to enhanced ion heating, reduced outflow speeds, and a modified reconnection geometry.
In contrast to previous Hall Magnetohydrodynaic (MHD) studies, we find that reconnection becomes a more efficient plasma heating mechanism in the presence of sub-Alfvénic flow shear, with ion heating increasing by as much as 300\%.
This enhanced heating is achieved by efficiently converting the incoming flow shear energy into thermal energy through istropization in the exhaust.
The enhanced heating leads to a pressure gradient away form the x-line exerting a force that reduces the outflow jet speed and slows down the reconnection process.
This conversion is due to beam selection effects, mixing and scattering in the exhaust. A theoretical model is developed which predicts well the exhaust heating and outflow speed reduction.
These results offer a potential explanation for recent Parker Solar Probe observations of suppressed reconnection in the presence of flow shear and carry significant implications for energy dissipation in turbulent plasmas.

\end{abstract}

\section{Introduction}
Magnetic reconnection is a plasma process that efficiently converts magnetic energy into plasma heating and Alfvénic outflows. This process is crucial in a variety of low-collisional plasma environments, spanning space physics and astrophysics. Initially proposed to explain solar flares, reconnection has since been repeatedly observed by satellites \textit{in situ} throughout the heliosphere, from the large-scale dynamics of the heliospheric current sheet and the Dungey cycle to the small-scale dissipation of energy in turbulence\citep{dungey61,paschmann+13,burch+16b,gosling+07a,gosling+07b,chasapis+18b,phan+20}.

Despite its importance and prevalence, our understanding of reconnection is primarily based on simplified models, which often assume constant and static inflowing plasma parameters. A critical complexity often ignored is the presence of a relative bulk flow across the reconnecting current sheet, i.e., \emph{flow shear}. An issue with this simplification is that reconnection inflow configurations with a flow shear are common in various systems, including current sheets in the solar wind of the inner heliosphere\citep{phan+20}, at the magnetopause flanks\citep{fuselier+05}, and within turbulent plasmas where they form at the boundaries of swirling eddies \citep{hasegawa+16,phan+96,matthaeus+86}.

Previous studies have examined flow shear reconnection with applications to the magnetopause; these studies used both Magnetohydrodynamics (MHD) and kinetic Particle-in-Cell (PIC) simulations to study the modifications to reconnection \citep{labelle-hamer+95,wang+08,wang+12,wang+15,doss+16,liang+24}.
Many of these studies focused on asymmetric reconnection, and/or a flow shear perpendicular to the reconnecting magnetic field. While applicable for parts of the magnetopause, none of these studies considered field-aligned flow shear for symmetric, kinetic systems, which is potentially more applicable to reconnection in collisionless plasma turbulence.
However, one such work considered the field-aligned symmetric case with Hall Magnetohydrodynamics (MHD) simulations which revealed several distinct modifications to the reconnection process, such as tilting of the exhaust region and a reduction in outflow jet speeds\citep{cassak+11,cassak11}. 
These studies presented Sweet-Parker-like scaling relations for the reduced outflow velocity; it was claimed that the kinetic energy associated with the flow shear would be responsible for decreasing the total available energy. 
However, these scaling predictions did not account for the fact that flow shear, in principle, should increase the total energy available in the system, suggesting that these relationships were empirical.

Consequently, our understanding of how field-aligned flow shear modifies magnetic reconnection remains incomplete, particularly from a kinetic perspective. Recent observations from Parker Solar Probe (PSP) highlight this knowledge gap, where reconnection appeared suppressed at relatively low, sub-Alfvénic flow shear values\citep{phan+20} - results at odds with the previous Hall-MHD and asymmetric PIC simulations\citep{cassak11,doss+16}. 

This manuscript investigates the effects of upstream, field-aligned flow shear on anti-parallel reconnection from a kinetic (collisionless) perspective.
While our results share some similarities with previous studies, we also show that reconnection becomes a significantly more efficient plasma heating mechanism in the presence of a flow shear.
This enhanced heating reduces the outflow jet speed and slows the reconnection process, which could explain recent PSP observations and significantly impact our understanding of reconnection as an energy dissipation mechanism in turbulent plasmas.

The manuscript is structured as follows: Sec.~\ref{sec:sims} details the code and simulations used in this study, and Sec.~\ref{sec:results} presents the results. Sec.~\ref{sec:theory} develops a physical argument for how flow shear modifies the reconnection process, leading to a prediction for the enhanced ion heating and the associated outflow reduction which is consistent with the simulation results. Finally, we discuss the potential implications of these results in Sec.~\ref{sec:discuss} and conclude in Sec.~\ref{sec:conclusion}.

\section{Simulations}\label{sec:sims}
To investigate the kinetic effects of velocity shear on magnetic reconnection, we perform a survey of simulations in 2.5D (2 dimensions in real space and 3 in velocity) using the Particle-In-Cell (PIC) code P3D \citep{zeiler+02}.
The simulation code is normalized to arbitrary density and magnetic field values of $B_0$ and $n_0$.
From this, the rest of the units are determined;
time to $t_0 = \Omega_{ci}^{-1} = m_ic/eB_0$, velocity to the Alfvén speed $v_0 = V_A =B_0/\sqrt{4\pi m_i n_0}$, 
lengths to the ion inertial length $l_0 = t_0v_0 = d_i = v_A/\Omega_{ci}$, electric fields to $E_0 = B_0 v_A/c$, where $c$ is the speed of light and temperatures\footnote{Note that in this work we take temperature in units of energy, i.e., $T$ has absorbed Boltzmann's constant $k_B$} to $T_0 = m_i V_A^2$.
Following standard PIC approaches, we use artificially reduced constants to reduce the computational cost of the simulations, including a reduced ion-to-electron mass ratio of $m_i = 25 m_e$ and speed of light $c/V_A = 7$.
We do not expect these choices will adversely impact the present study.
The ions are initialized with an ion plasma beta of $\beta_i = 2T_i/m_i V_A^2 = 0.1$, and electrons with $T_e/T_i = 1$.

The simulations use a regular Cartesian grid with doubly periodic boundary conditions, with lengths $L_x \times L_y = 81.92 d_i \times 40.96 d_i$, with grid scale $\Delta x = d_i/50$, time step $\Delta t = (200 \Omega_{ci})^{-1}$ and with 500 macro-particles per cell.
We initialize the simulation with a double-tanh current sheet\footnote{This is not precisely a Harris current sheet because both the background and enhanced density population in the current sheet carry the current, rather than just the extra particles.}, 
with the initial reconnecting magnetic field given by $B_x/B_0 
= \tanh ((y -0.25 L_y)/w_0)
-\tanh [(y-0.75L_y)/w_0]-1$, 
where $w_0 = d_i/2$ is the half current sheet thickness.
The temperature is held constant across the sheet and the density is increased to insure pressure balance.
The ions and electrons are initialized with drifting Maxwellian distributions with a bulk velocity shear parallel to the magnetic field, i.e., $u_x = V_{\rm shear} B_x/B_0$, where $V_{\rm shear}$ is the amplitude of the flow and is the independent variable changed between simulations, and kept between $0 - V_A$ as to remain stable to the Kelvin-Helmholtz instability\citep{chandrasekhar61}.
We limit the scope of this paper only to consider antiparallel reconnection, i.e., where $B_z = 0$ initially.
We trigger reconnection by adding a small amplitude perturbation in $B_y$ of the form $B_y/B_0 = \psi_0 \sin^{10}{(x/L_x)}\cos{(x/L_x)}$ where $\psi_0 \ll 1$, an approach adopted in several previous studies on reconnection\citep{haggerty+18,mbarek+22,giai+24}. 

In this work, six simulations have been performed with increasing values of flow shear $\vs$ from $\vs = 0 - 1V_A$ in increments of $V_A/5$. The simulations are run until the islands generated by the exhaust reach the simulation edges, allowing for steady-state reconnection to be reached and analyzed before this occurred.

\section{Results}\label{sec:results}
\subsection{Overview}
\begin{figure*}[ht]
    \centering
    \includegraphics[width=0.95\textwidth]{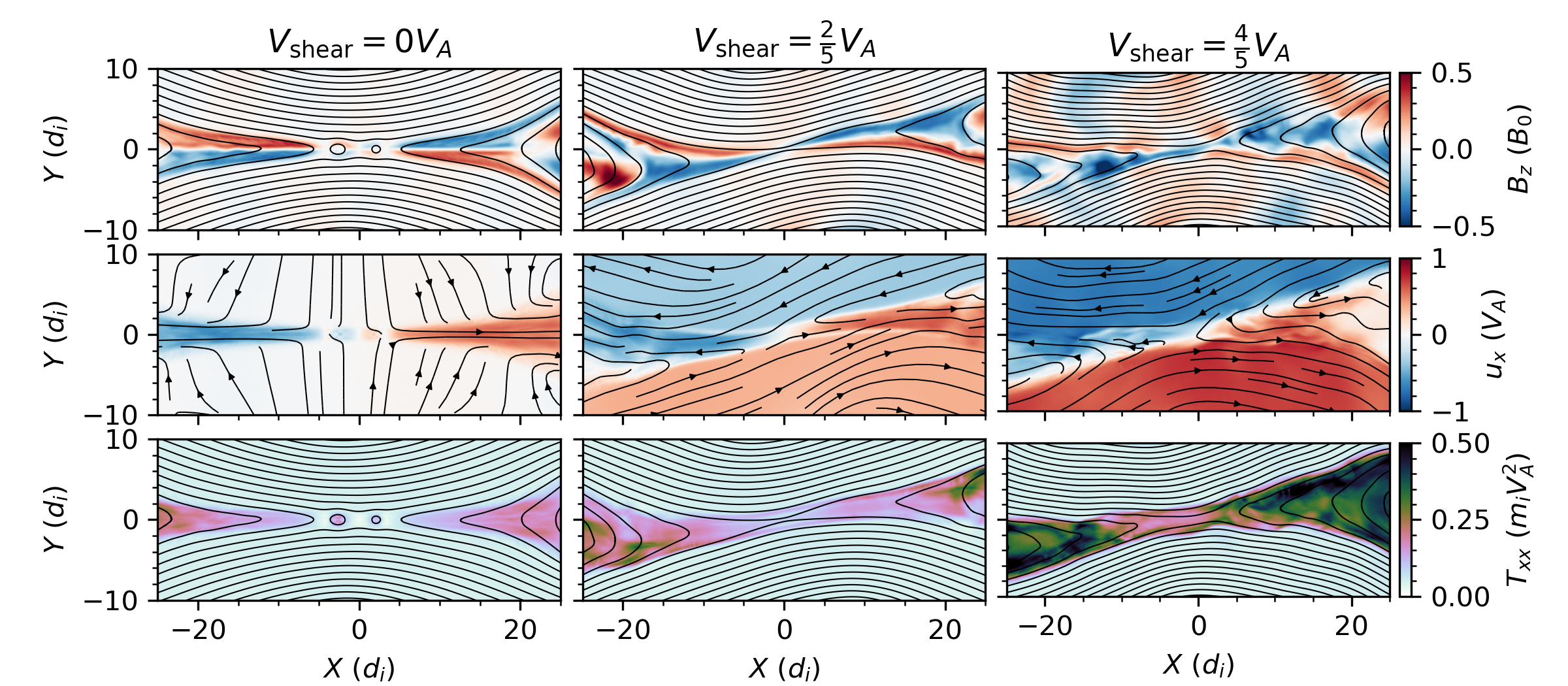}
    \caption{2D snapshots from different simulations: the out of plane magnetic field $B_z$ (Top Row), the ion bulk flow in the outflow/$x$-direction $u_x$ (Middle Row), and the $xx$-element of the ion temperature tensor $T_{ixx}$ (Bottom Row). 
    From left to right, the simulations were initialized with sear flows of $0\, V_A$, $(2/5)\, V_A$ and $(4/5)\, V_A$. This overview plot shows the key effects of increasing flow shear on reconnection including the tilting of the current sheet, the reduction of the outflow velocity and the significant enhancement in ion exhaust heating.}
    \label{fig:overview}
\end{figure*}
The simulations show that many of the hallmarks of reconnection remain, even with an increasing upstream velocity shear. Fig.~\ref{fig:overview} shows 2D images from 3 different simulations ($\vs/V_A = 0, 2/5, 4/5$ in columns from left to right) for three different variables (the out-of-plane magnetic field, $B_z$ (Top Row), the bulk outflow velocity along the reconnection field direction, $u_{x}$ (Middle Row), and the $xx$-element of the ion pressure tensor $T_{ixx}$ (Bottom Row). Each row is plotted with the same color limits for direct comparison. Even as $\vs$ increases, the out-of-plane magnetic field shows the quadrupolar structure associated with Hall physics \citep{shay+01}. The outflow jets and associated ion heating persist in the presence of a flow shear, as shown in the exhaust regions in the 2 right-most rows.

However,as $\vs$ is increased, we also find differences from shear-free reconnection.
Notably, the reconnection exhaust is rotated counter-clockwise (rotation vector along $\hat{z}$), with the degree of the tilt increasing with $\vs$, as is evident from the second and third columns of Fig.~\ref{fig:overview}.
More precisely, the exhaust is tilted towards the side where the outflow jet is opposite the flow shear.
This can be seen from the black flow lines plotted over the top of $u_x$ in the middle row.
This tilting was also identified in Hall-MHD simulations of shear-flow reconnection\citep{cassak11}.
In this work, the tilt was attributed to a difference in the gradient of the dynamic pressure in the $x$-direction, contributing to a right-handed, out-of-plane torque.

Additionally, we find that the bulk outflow velocity is reduced for increasing values of $\vs$, as demonstrated in the second row of Fig.~\ref{fig:overview};
this result is consistent with previous Hall-MHD studies as well.
However, the bulk outflow speed in all simulations is sub-Alfvénic, which differs from the fluid results in which the outflow speed was equal to the Alfvén speed for the shear-free fluid simulations (see Fig.~4 from \citet{cassak11}).
The reduced outflow speed in kinetic simulations was discussed in \citep{haggerty+17,li+21,giai+24} and attributed to increased ion thermal pressure in the exhaust for the $xx$-element.
Furthermore, because all outflow speeds are reduced in the kinetic regime, the outflow speeds will become less than the flow shear velocity for relatively modest flow shear values (this is estimated to occur around $\vs \gtrsim V_A/2$).
Above this flow shear threshold, it may become difficult to distinguish a reconnection outflow jet from the bulk motion of the upstream plasma.
This threshold could be connected to the apparent suppression of reconnection in the presence of a modest flow shear as identified with PSP in the solar wind\citep{phan+20}.

Finally, the most dramatic effect of increasing flow shear speed is the increase in the ion exhaust temperature.
This is shown for $T_{ixx}$ in the bottom row of Fig.~\ref{fig:overview}.
As $\vs$ increases, the ion temperature increases dramatically by as much as 300\% for the largest shear speed simulation.
This effect was not identified in previous studies and runs counter to current scaling predictions for ion heating in reconnection, with most studies predicting that the ion heating will be proportional to the outflow speed squared, $\Delta T_i \propto m_i u_x^2$ \citep{drake+12,shay+14,haggerty+15,haggerty+16b,haggerty+18,phan+13,phan+14,yoo+14}.
While it is true that increasing the upstream Alfvén speed increases both the outflow speed and the ion heating, the figure shows that for a fixed upstream Alfvén speed, $|u_x|$ is actually anti-correlated to $\Delta T_i$, in the presence of a flow shear.
Notably, each of these simulations has the same magnetic energy density per particle, which means that for the same magnetic field strength, reconnection with a flow shear results in more internal energy.

\begin{figure}[htbp!]
    \includegraphics[width=0.45\textwidth]{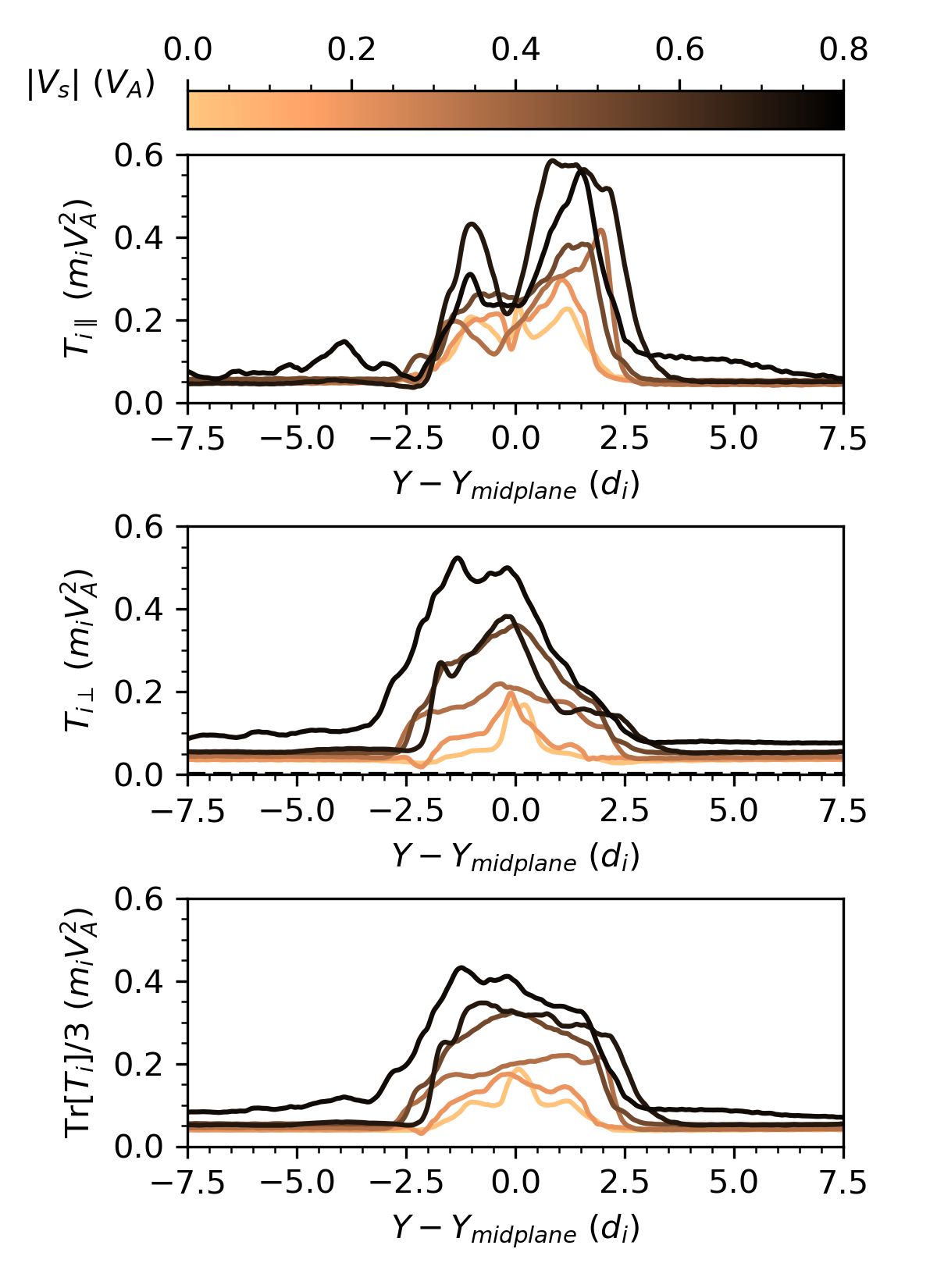}
    \caption{Cuts along $y$ of the ion temperature tensor rotated into the frame of the local magnetic field. The cuts are taken $15\,d_i$ downstream of the x-line in each simulation after steady state reconnection has been reached. Parallel $T_{i\parallel}$ (Top Row), perpendicular $T_{i\perp}$ (Middle Row) and average ion temperature $(T_{i\parallel} + 2 T_{i\perp})/3$ (Bottom Row)for simulations with different measured flow shear speeds. The color corresponds to the measured upstream flow speed as indicated by the colorbar. The ion temperature in the reconnection exhaust increases with the inflowing flow shear speed.}
    \label{fig:temps}
\end{figure}

The ion heating enhancement is further demonstrated in Fig.~\ref{fig:temps}, which shows cuts across the exhaust for the six different simulations for the field-aligned, perpendicular, and the effective ion temperature (trace of the ion temperature tensor divided by 3) (from top to bottom).
The color of each line corresponds to the upstream flow shear speed in the simulation measured at the time when the cut is made, as denoted by the color bar (note that the upstream flow shear speed for $\vs = V_A$ simulation decreases as the simulation evolves).
The cuts are taken $15\ d_i$ downstream of the x-line and centered about the midplane.
Each measure of the ion temperature increases substantially; however this increase is not strictly isotropic, as both $T_{i\parallel}$ and $T_{i\perp}$ vary along the $y$-direction.
Finally, we note that the electron heating (not shown) was not found to appreciably affected by increasing flow shear speed in the simulations.

\subsection{Empirical Scaling}
\begin{figure}[ht]
    \centering
    \includegraphics[width=0.45\textwidth]{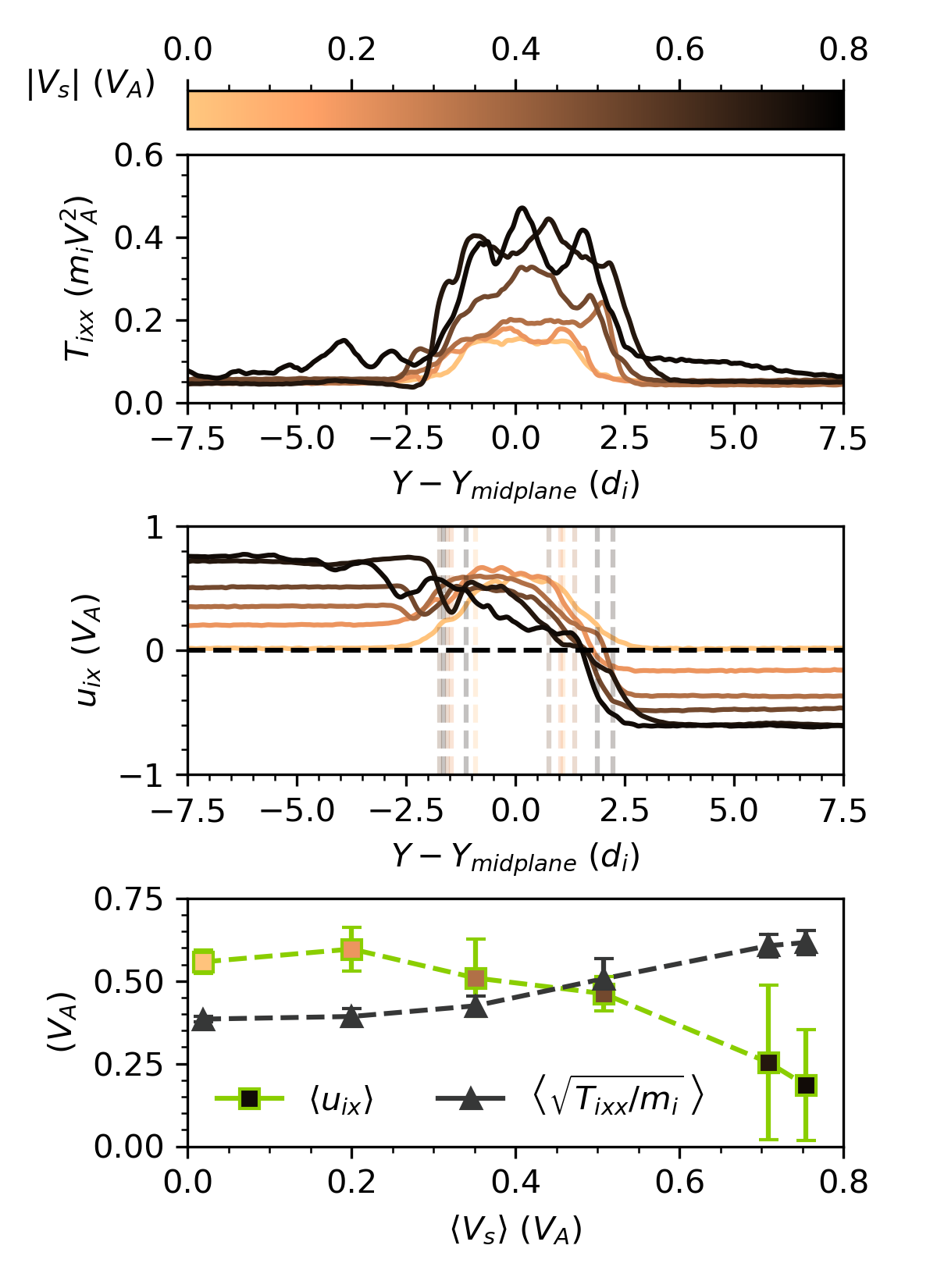}
    \caption{Cuts of the $xx$-element of the ion temperature tensor (top) and the $x$-element of the ion bulk flow vector (middle).
    The plots use the same conventions as Fig.~\ref{fig:temps} and are taken at the same locations and times.
    The light dashed color-coded lines show the locations that the averages were determined.
    The bottom shows the average values measured within the exhaust as a function of upstream flow shear speed for the bulk outflow (green outlined squares) and the ion thermal speed based on the $xx$-element of the ion temperature tensor (brown outlined triangles).}
    \label{fig:averages}
\end{figure}

We further explore the effect of flow shear by determining empirical scaling relationships.
We do this by determining various average values in each of the simulations.
For a given cut along the $y$-direction through the exhaust, the shear speed is averaged several $d_i$ upstream of the reconnection exhaust, and the outflow and ion heating are averaged in the exhaust.
The averaging limits are chosen manually for each simulation and were selected based on where the ion temperature increases and the shear flow has a rapid spatial change, roughly corresponding to the separatrix layer.
Examples of these cuts are shown for $u_x$ and $T_{ixx}$ in the first and second panels of Fig.~\ref{fig:averages}, respectively, along with the averaging areas shown by lower opacity dashed lines\footnote{We average $T_{ixx}$, as it will ultimately be the only element of the ion temperature tensor which is used to determine a prediction for the outflow speed and heating.}.
The colors of each cut correspond to the average upstream flow shear speed, similar to Fig.~\ref{fig:temps}.
The top panel shows that the ion exhaust temperature increases with the flow shear speed, even while the outflow speed decreases, as shown in the second panel.

The average values are plotted as a function of flow shear speed in the bottom panel of Fig.~\ref{fig:averages}.
The green squares show the outflow speeds $\left < u_x \right >$ averaged from the simulations, where the faces have been color-coded according to the upstream flow shear value.
In contrast, the brown triangles show the square root of the average ion temperature\footnote{This was chosen for direct, dimensional comparison with the outflow velocity} $\sqrt{\left < T_{ixx}/m_i \right >}$.
The error bars are determined from the standard deviation over the averaging area.
The figure further demonstrates that the two quantities are directly anti-correlated, with the outflow dropping from $\sim 0.62 V_A$ to $0.17 V_A$, roughly a $\sim 70\%$ reduction in the outflow speed and a $\sim 90\%$ reduction in the kinetic energy of the bulk flow; 
simultaneously, the increase in the $xx$-element of the temperature tensor goes from $\Delta T_{ixx} \sim  0.11 m_i V_A^2 $\footnote{Note that this value is consistent with the heating values found in both simulations\citep{haggerty+15} and observations \citep{phan+14}} to $\sim 0.35 m_i V_A^2$, corresponding to over a $300\%$ increase in the ion heating.
Note that the term being subtracted off is the inflowing upstream temperature.
Following the approach in \citet{shay+14} and normalizing the change in ion enthalpy density to the inflowing Poynting flux, $R_H$, we find that $R_H \approx \frac{\gamma}{\gamma - 1} \Delta T_{ixx}/m_i V_A^2 \sim 0.88$, or about $90\%$ of the available magnetic energy per particle.
Note, this estimate estimate assumes the exhaust temperature is isotropic, implying that $\gamma = 5/3$ would be the adiabatic index, which is roughly supported by Fig.~\ref{fig:temps} .
While it is tempting to interpret this quantity as the fraction of magnetic energy that goes into heating, the analysis of \citet{shay+14} did not include the inflowing energy associated with the flow shear; in fact, the inflowing flow shear energy is being efficiently converted into thermal ion energy, as will be demonstrated later in the manuscript.
However, for this simulation the relatively large flow shear speed of $4/5V_A$ can only account for $1/2m_i(4/5V_A)^2 \sim 0.32m_i V_A^2$ of the thermal enhancement, leaving $\sim 0.56 m_i V_A^2$ of the heating associated with the release of magnetic energy, which is still a factor of 2 larger than the normalized change in the enthalpy in the shear-flow-free case.

\section{Theory}\label{sec:theory}
The dramatic increase in ion exhaust temperature is a kinetic effect of antiparallel (and nearly antiparallel) reconnection.
For the inflow side with a flow shear directed towards the x-line (i.e., upper right and lower left quadrants of the rightmost column in Fig.~\ref{fig:overview}), the plasma enters the exhaust and reaches the midplane, where the field line is sharply kinked. 
At this location, ions with even a modest velocity will demagnetize and pitch angle scatter, directly transferring their bulk flow energy into thermal energy.
Demagnetization is expected to occur when the current sheet width becomes comparable to the ion gyroradius based on the bulk shear flow velocity at the midplane, $r_{g,s} = \vs m_i c/q B_y \sim d_i \rightarrow \vs \sim \frac{1}{10}V_A$, where the field strength of the midplane is taken to be one-tenth of the upstream reconnecting field, corresponding to a reconnection rate of 0.1\citep{speiser65,birn+01}. The physics of ion scattering are fairly complex and will be covered in detail in a forthcoming companion paper, along with the corresponding electron dynamics. In this manuscript, we focus on the effect of this enhanced ion heating on the reconnection process.

The reduction in the ion outflow speed is connected to the enhanced ion heating by considering force balance along the outflow direction in the exhaust.
This approach has been used previously to explain the reduction in the outflow speed in reconnection in both antiparallel \citep{haggerty+17,li+21} and guide field \citep{giai+24} configurations. 
As reconnection heats the outflowing plasma, a pressure gradient develops between the exhaust and the x-line.
The pressure gradient exerts a force against the outflow direction, opposing the magnetic tension force and ultimately reducing the outflow speed to sub-Alfvénic values.
An equilibrium is reached as the heating depends on the magnitude of the outflow speed, as discussed/demonstrated in many previous studies in both simulations (e.g., \citet{drake+12,shay+14,haggerty+15,haggerty+16b,haggerty+18}) and observations (e.g., \citet{phan+13,phan+14,yoo+14,oieroset+24}). 
Thus an expression for the scaling of the ion temperature will be a function of both the outflow speed $V_o$ and the flow shear speed $\vs$.

\begin{figure*}[htbp!]
    \centering
    \includegraphics[width=0.95\textwidth]{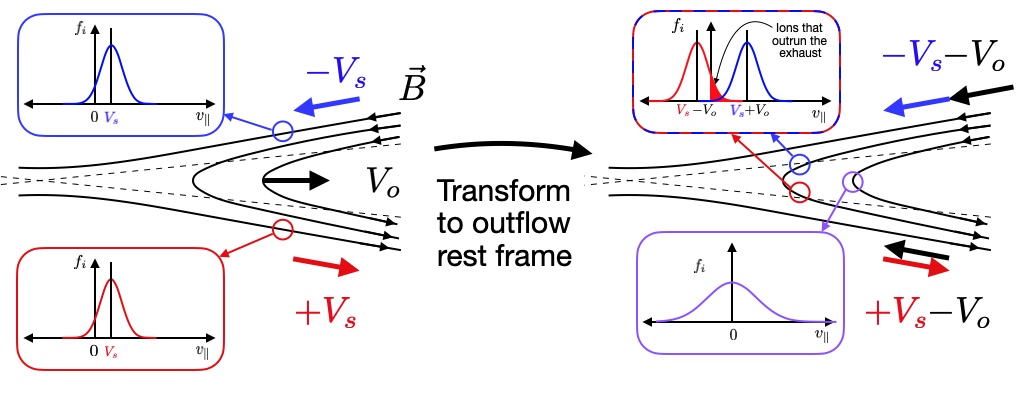}
    \caption{Diagram of how the exhaust temperature is determined from the inflowing ion populations.
    The blue distribution function represents ions with a flow shear opposite that of the outflow velocity, while the red distribution function shows ions with a flow shear in the same direction, however both populations are parallel to the local magnetic field, as shown in the corresponding distribution functions on the left.
    The distributions are boosted into the outflow frame as shown on the right side of the diagram. The composite, counter-steaming distribution is shown in the upper box, with the shaded red region showing ions that never reach the exhaust because their parallel speed is greater than the outflow speed. The two distribution functions are taken to isotropize and produce a Maxwell-Boltzmann distribution at rest in this frame with the same kinetic energy as the combined inflowing distribution functions, as shown in the purple box in the bottom right.}
    \label{fig:diagram}
\end{figure*}

We derive a prediction for the ion heating by considering the fraction of inflowing ions that reach the exhaust and scatter.
This calculation is most easily performed in the reference frame moving with the bulk outflow velocity.
We take the outflow velocity vector to point in the $+\hat{x}$ direction, where the magnetic field and flow shear direction are along $+\hat{x}$ below the exhaust and $-\hat{x}$ above the exhaust; with this configureation there is a positive (anti-clockwise) tilt around the $+\hat{z}$ axis, similar to the exhaust on the right-hand side of the x-line in the 3rd column of Fig.~\ref{fig:overview}.
From the simulations we find the current sheet tilting by $\theta \sim 10^\circ$ at most, which allows us to take the outflow as approximtly in the x-direction as $\cos(\theta) \gtrsim 0.98$.
Transforming into the reference frame moving with the outflow velocity ($\vo \hat{x}$) decreases the flow shear speed below the exhaust while increasing the flow shear speed above the exhaust.
In this frame, the lower population flows into the exhaust antiparallel to the field line, while the upper population flows away from the exhaust parallel to the field line;
because of this, the flow shear reduces the speed for the lower population while boosting the upper population.
Note that for a sufficiency large flow shear, $\vs > \vo$, half of the inflow population may be unable to enter the exhaust. 
A diagram of this is shown in Fig.~\ref{fig:diagram}, with the red and blue distribution functions corresponding to the the lower and upper sides of the exhaust, respectively.
The particles from the two distribution functions isotropize in the outflow rest frame (purple distribution on the right side of the diagram) when they reach the current sheet's center and produce a composite distribution function, shown in cyan. In this co-moving frame, there is no inductive electric field and so a particle's total energy is conserved; 
from this the total ion temperature of the isotropized distribution is the same as the average energy of the unscattered distribution function. 
The temperature of each diagonal element of the temperature tensor is then taken to be $T_{ixx} = T_{iyy} = T_{izz} = {\rm Tr}[T_i]/3$.

The calculation, however, is complicated as the entirety of both inflowing ion distributions do not reach the exhaust.
Some fraction of the inflowing ions have a parallel velocity away from the exhaust large enough to `outrun' the outflow, i.e., the ions still have a velocity in the positive x-direction, even in the outflow frame. 
The ions which do not enter the exhaust is set by both the thermal speed of the upstream ions and the flow shear velocity relative to the speed of the outflow jet.
Since the outflow speed is sub-Alfvénic, even a sub-Alfvénic flow shear will result in a potentially significant fraction of the ions outrunning the exhaust.
Note that the effect of the flow shear is asymmetric, as ions on the lower half of the current sheet are moving away from the current sheet, while ions on the upper half are moving towards the current sheet.
The portion of the ion population that outruns the exhaust is denoted by the red-shaded region in Fig.~\ref{fig:diagram}. 
An interesting side effect of this is that for increasing flow shear, the exhaust is increasingly populated by ions from one-half of the exhaust; this effect is examined in more detail and demonstrated in PIC simulations in the forthcoming companion paper. 

\begin{figure}[htbp!]
    \includegraphics[width=0.45\textwidth]{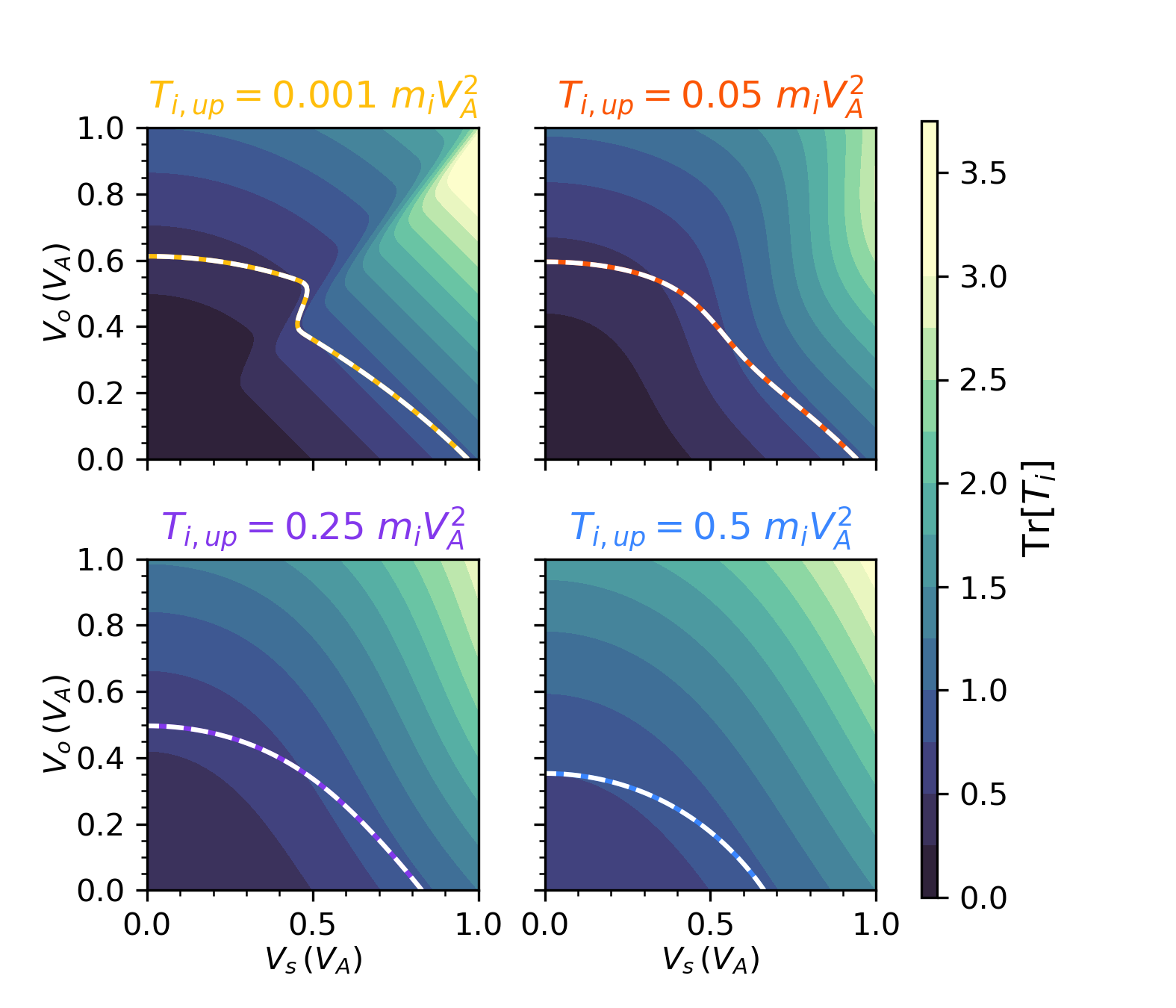}
    \caption{2D maps of the predicted ion exhaust temperature as a function of outflow speed ($V_o$, $y$-axis) and flow shear speed ($V_S$, $x$-axis), for different inflowing ion temperatures as labeled in the title of each subplot. The white dashed lines correspond to the pressure ballance/outflow constraint being satisfied (i.e., Eq.~\ref{eq:contraint}). The dashed white and black line provides the prediction for the outflow as a function of flow shear speed.}
    \label{fig:predict}
\end{figure}

Combining these two effects yields a relationship between the total ion exhaust temperature and both the outflow and flow shear speeds.
An analytical form can be determined which includes the error function ${\rm erf()}$ function, however the expression is so long as to be impractical to use.
Alternatively we elect to solve for the temperature directly by numerically integrating the analytical, composite distribution functions. 
Fig.~\ref{fig:predict} shows the predicted total ion exhaust temperature as a function of both the upstream flow shear speed (horizontal axis) and the outflow speed (vertical axis) for four different upstream ion temperatures ($T_{i,up} = 0.001, 0.05, 0.25, 0.5 m_iV_A^2$ for the top left, top right, bottom left and bottom right subplots, respectively). 
The increased heating along the vertical direction corresponds to reconnection more efficiently heating inflowing plasma through the contraction of the magnetic field line, while the increased heating along the horizontal direction corresponds to more flow shear energy being channeled into heating. 

We determine an ion exhaust temperature for a given outflow and flow shear speed using another relationship to close the equation.
This comes from balancing the forces along the x-direction in the reconnection exhaust, following the approach in \citet{Liu+17}, \citet{li+21} and \citet{giai+24}.
Force balance yields
\begin{align}
\frac{1}{2}m_i n_o V_o^2 + \frac{B_y^2}{8\pi} + \Delta P_{xx} = \frac{L}{\delta} \frac{\varepsilon B_y B_x}{8\pi},\label{eq:contraint}
\end{align}
where $m_i$ is the ion mass, $n_o$, $V_o$ and $B_y$ are the number density, outflow velocity and the normal magnetic field measured in the exhaust, $B_x$ is the reconnecting magnetic field measured upstream of the ion diffusion region, $\Delta P_{xx}$ is the change in the $xx$-element of the pressure between the x-line and the exhaust, $L/\delta$ is the aspect ratio of the ion diffusion region, and $\varepsilon$ is the firehose parameter of the inflowing plasma $\varepsilon  \equiv 1 - 4\pi(P_\parallel - P_\perp)/B^2$.
This equation implies that the force due to magnetic tension (right hand side of the equation) is reduced by an increasing $xx$-pressure gradient along the outflow direction ($\Delta P_{xx}$);
this pressure gradient develops as the ions are heated in the exhaust relative to those in the diffusion region, as shown in Fig.~\ref{fig:overview}. 
This relationship is simplified by assuming a fast reconnection rate (i.e., $B_y/B_x \approx \delta/L \sim 0.1$) and assuming that the density is constant between the exhaust and diffusion region,
\begin{align}
\frac{n_o}{n_0}\left ( V_o^2  + 2\frac{\Delta T_{ixx}}{m_i} \right ) \approx \varepsilon V_A^2,
\end{align}
where $n_0$ is the upstream number density.
This equation serves as a constraint for the outflow-shear-heating relationship.
Note we have neglected the contribution from electron heating as it is expected to be five times smaller than the change in ion heating\citep{phan+13,shay+14,phan+14,haggerty+15}. 
For the simulations in this study and for the rest of the paper we take the upstream temperature to be isotropic and $\varepsilon\rightarrow 1$
The contour defined by this constraint is shown by the dashed white lines in Fig.~\ref{fig:predict}, where we have used an exhaust density of $n_o = 1.6 n_0$ consistent with what is approximately found in the simulations
\footnote{An alternative approach to solve these equations is to use the density inferred from the distribution functions shown in Fig.~\ref{fig:diagram}, however in this approach we have found that the density is over predicted for low shear and under predicted for stronger shear. The exhaust density should be set by a combination of parallel inflow and perpendicular inward drifts that account for the tilting of the exhaust and pressure balance. This simplified analysis of selecting a single exhaust density still yields results consistent with simulations}. 
The white dashed line shows for a given flow shear speed, $\vs$, what outflow speed, $\vo$, is needed to produce the ion heating so that Eq.~\ref{eq:contraint} is satisfied, which demonstrates the predicted relationship between $\vs$ and $\vo$.

\begin{figure}[htbp!]
    \includegraphics[width=0.45\textwidth]{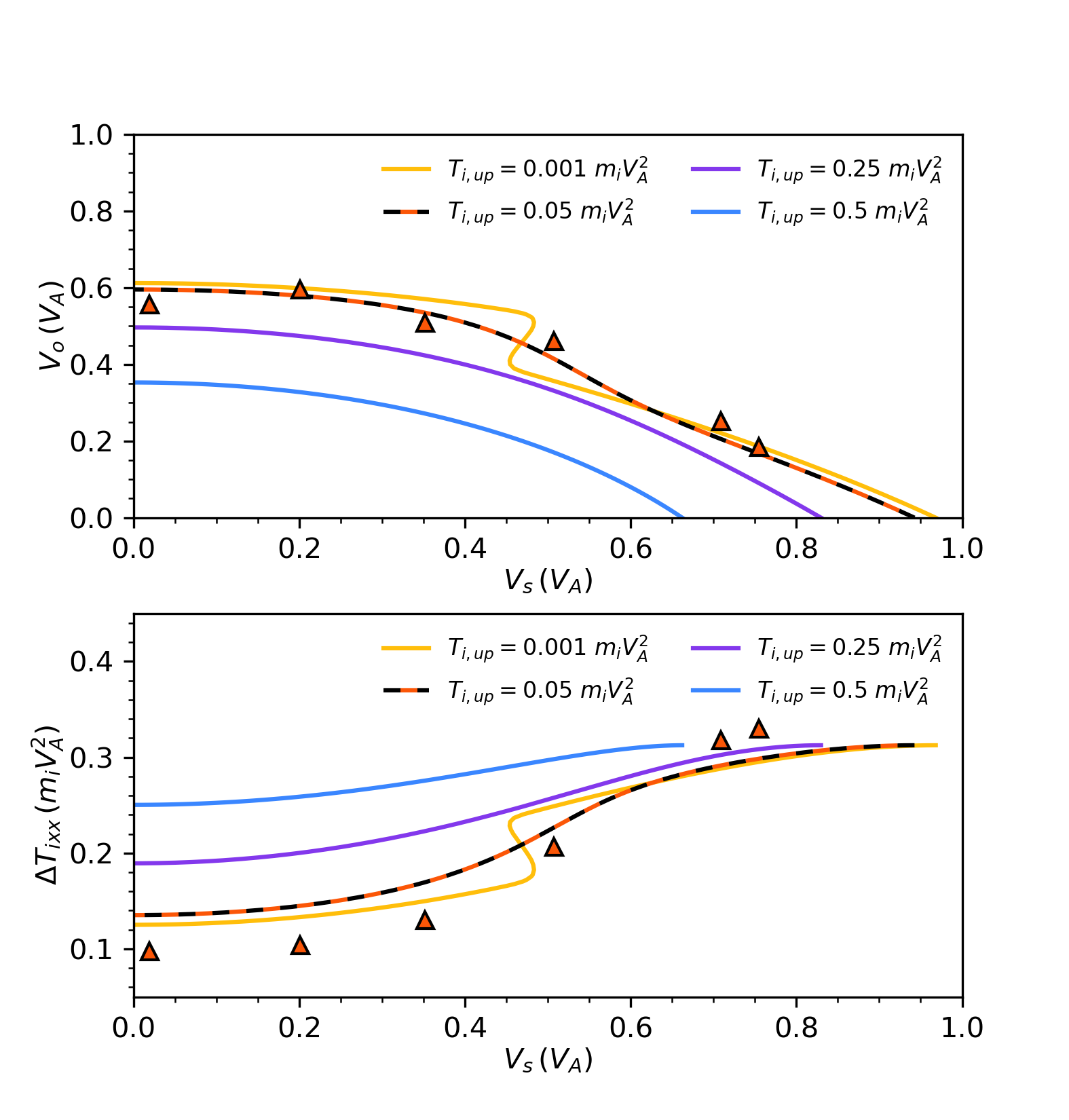}
    \caption{Comparison between simulations (orange triangles) and predictions (multi-colored solid lines) for the outflow velocity (top panel) and ion heating (bottom panel) as a function of inflowing flow shear speed. The different colors correspond to the predictions for different initial ion temperatures, as described in the legend, with the black  dashed, darker orange having the same initial temperature as the simulations. }
    \label{fig:predict_sims}
\end{figure}

Using this constraint, we can find a solution for both the outflow speed and the ion heating as a function of upstream flow shear, which is shown in Fig.~\ref{fig:predict_sims} for different values of upstream $T_i$ (or equivalently $\beta_i$ for fixed Alfvén speed).
The top panel shows the predicted outflow speed $V_o$ as a function of flow shear.
These predictions are the same as the contours in Fig.~\ref{fig:predict} and color-coded in the same manner (based on inflowing temperature).
The contours show that as the flow shear speed increases the outflow speed decreases consistent with what was found in simulations.
The orange triangles show the average values determined in the simulation initialized with an upstream ion temperature of $0.05 m_i V_A^2$.
The prediction is in excellent agreement with the corresponding simulations, in both the magnitude of the values and the trend as a function of flow shear speed.
Note that the sub-Alfvénic outflow speeds are consistent with the thermal effects discussed in \citet{haggerty+18}, \citet{li+21} and \citet{giai+24} and are required to predict both the outflow speed and ion exhaust heating accurately.

The bottom panel of Fig.~\ref{fig:predict_sims} shows the heating as a function of flow shear using the same approach.
The ion exhaust heating increases with increasing flow shear, as was found in the simulations.
We find a factor of 3 increase for the heating in the strong shear limit relative to the shear-free case.
Again, the average heating in the simulations agrees with the corresponding (orange) prediction;
however, the heating is slightly overpredicted for small flow shear speeds and underpredicted for larger flow shear speeds.
This slight discrepancy is likely linked to the simplifying assumptions about the exhaust density.
These two plots show that the larger flow shear speed increases the ion exhaust heating, but this heating increase comes at the expense of the bulk outflow speed. 
The outflow speed reduction directly results from the back pressure gradient opposing the magnetic tension force, and all three terms counterbalance.

It is noteworthy that the very low-temperature limit has an ``S'' shaped feature in the yellow line for both the top and bottom of Fig.~\ref{fig:predict_sims}. The doubling-back of the contour near $\vs \sim 0.5 V_A$ corresponds to the point where $\vs > \vo$. In the low-temperature limit, this corresponds to one of the inflowing populations never reaching the exhaust.
We suggest that the temperature would agree with either the upper or lower solution based on the history of the outflow speed;
if the flow shear is constant and the exhaust starts from rest, one might expect the outflow speed to take the smaller value and the temperature to take the larger value.
This feature likely results from the overly simplified model and is unlikely to manifest in a physical system; further examination is beyond the scope of this manuscript.

These predictions were determined by integrating different distribution functions and then numerically inverting the relationship, and as such we recognize that carrying out this approach for comparison/prediction with future simulations and observations would prove arduous.
Because of this we provide a relatively simple, normalized equation that reasonably well-approximates the predictions shown in Fig.~\ref{fig:predict_sims},
\begin{equation}
\frac{V_o}{V_A} \approx C_0  - C_1  \frac{V_s^2}{V_A^2} - C_2\left( \beta_i +  \frac{\beta_i^2}{2}\right ) - C_3 \frac{V_s}{V_A} \beta_i + C_4 \frac{V_s}{V_A},\label{eq:estimate}
\end{equation}
where $C_0 = 0.62$, $C_1 = 0.75$,  $C_2 = 0.155$, $C_3 = 0.06$, $C_4 = 0.07$.
This approximation is determined with a 2D, quadratic regression, and as such it does not accurately represent  the ``S'' shape for very low beta plasmas previously discussed.
Despite this, the equation agrees well for much of parameter space, providing a simple approximation for the outflow speed in the shear-free case, as well as the maximum flow shear as a function of inflowing ion plasma beta.
Furthermore, we present a prediction for the ion heating using this approximation and the pressure balance constraint from Eq.~\ref{eq:contraint},
\begin{equation}
    \frac{\Delta T_{ixx}}{m_iV_A^2} \approx \frac{5}{16} - \frac{1}{2}\left ( \frac{V_o}{V_A} \right )^2,
\end{equation}
where $\vo/V_A$ can be determined from upstream quantities using Eq.~\ref{eq:estimate}

\section{Discussion}\label{sec:discuss}
The results presented in this manuscript are likely impactful for disparate, astrophysical plasma systems where reconnection occurs;
perhaps most immediately for the dissipation in collisionless plasma turbulence.
An outstanding problem in plasma turbulence is determining the physical mechanisms responsible for dissipating energy in turbulence when collisions are infrequent\citep{matthaeus+20}, and magnetic reconnection has been suggested as a potentially important candidate.
Reconnection has consistently been shown to occur at current sheets formed by turbulence (e.g., \citep{retino+07,gosling+07a,servidio+09,wan+13,haggerty+17,shay+18,chasapis+18b,zhou+21,ergun+22,stawarz+22}).
These current sheets form at the boundary of swirling eddies which are likely to have a relative flow shear.
Because of this, most reconnection events in turbulence would likely include a flow shear component, and so the results of this manuscript suggest that reconnection may be even more of an effective dissipation turbulent dissipation mechanism than previously thought.

However, in this work we have asserted that flow shear energy, rather than magnetic energy, is converted into heating.
This implies that magnetic reconnection may more efficiently dissipate flow energy rather than magnetic energy depending on the boundary layer/current sheet conditions.
This could represent a paradigm shift in the role of reconnection for energy dissipation in turbulence, where reconnection changes the magnetic topology, which allows flow energy to be efficiently dissipated.
An obvious extension of this work is to identify reconnection in kinetic turbulence simulations and determine if the shear-flow enhanced heating rate would account for a significant fraction of the energy dissipation at these scales. 
While this is an important question to explore, we leave it to future work.

Beyond turbulence these results could impact our understanding of reconnection in the inner heliosphere. \citet{phan+20} reported observations of current sheets encountered by Parker Solar Probe (PSP) during its first approach to the Sun.
They identified numerous current sheets in which reconnection was or was not occurring based on qualitative determination of the presence or absence of outflow jets. It was found that current sheets with outflow signatures had velocity shear $< 0.6 \Delta V_A$, whereas current sheets without outflow signatures, deemed non-reconnection events, tended to occur in current sheets associated with magnetic switchbacks, where the tangential velocity shears across the current sheets were between $ 0.6 \Delta V_A$ and $1 \Delta V_A$.  \citet{phan+20} suggested that even a modest (sub-Alfvénic) velocity could already suppress reconnection. However, the results of the present manuscript suggest that the identification of reconnection outflows (described in \citet{phan+20} and \citet{gosling+05}) in the presence of flow shears may not be correct, since our simulations show that, even with a modest velocity shear of $0.7 \Delta V_A $, the velocity profile across the reconnecting current sheet does have the usually enhanced flow signature, even though reconnection is active. We thus suggest a re-examination of solar wind current sheets with large velocity shears to re-assess the occurrence of reconnection, which should not be based solely on the velocity profile. Strong ion heating might be a more reliable signature of reconnection events in large-shear-flow current sheets. The findings of such a study could have important implications for assessing the role of reconnection in the acceleration and heating of the near-Sun Alfvénic solar wind. 

While the results in this manuscript are potentially impactful for a range of space and astrophysical systems, more work must be carried out to determine the extent of this impact.
In particular, this work is limited to 2D, antiparallel reconnection with a low initial ion beta.
Reconnection occurring in turbulence will likely include a guide field, which may affect the results.
Furthermore turbulence is an inherently three-dimensional process, and the reduced dimensionality of these simulations could also impact the results.
While we presented a theory to include increasing ion beta, we have not verified these predictions with simulations.
Finally, this work has only considered the heating and feedback of the ion population.
However, electron physics may also be important, and even if reconnection is suppressed on the ion scales, it is unlikely to affect electron-only reconnection\citep{phan+18}.
These issues merit further study and discussion, and will be the focus of a future manuscript.

\section{Conclusion}\label{sec:conclusion}
In this work we show that the presence of a flow shear significantly alters the kinetic physics of magnetic reconnection in an anti-parallel configuration.
We demonstrate that magnetic reconnection is significantly affected by increasing the magnetic field aligned flow shear speed.
We reaffirm the reconnection geometry is titled and the outflow jet speeds are reduced, we show that the ions are heated more and derive a prediction for the outflow speed from first-principles using this enhanced ion heating. 
The increased ion heating is the most surprising and potentially consequential of these effects. While previous theories predicted that the ion heating should decrease with the reduced outflow speed\citep{drake+12,haggerty+15,haggerty+16b}, the simulations and theory presented in this work show the opposite effects. We find that for sub-Alf\'enic flow shear speeds, ion heating can increase by as much as 300\% by converting flow shear energy directly into thermal energy.
These results are potentially impactful for energy dissipation in plasma turbulence and for our understanding of heliospheric \emph{in situ} observations of reconnection.

\begin{acknowledgments}
The work of CCH was supported by the NSF/DOE Grant PHY-2205991, NSF-FDSS Grant AGS-1936393, NSF-CAREER Grant AGS-2338131; 
MAS by NASA grant 80NSSC20K1813 and 80NSSC24K0559
PAC by DoE grant DE-SC0020294 and NASA grant 80NSSC24K0172
Simulations were performed on TACC’s Stampede 2 and Purdue’s ANVIL, with allocations through NSF-ACCESS (formally XSEDE) PHY220089 and AST180008.\end{acknowledgments}


\begin{thebibliography}{}
\expandafter\ifx\csname natexlab\endcsname\relax\def\natexlab#1{#1}\fi
\providecommand{\url}[1]{\href{#1}{#1}}
\providecommand{\dodoi}[1]{doi:~\href{http://doi.org/#1}{\nolinkurl{#1}}}
\providecommand{\doeprint}[1]{\href{http://ascl.net/#1}{\nolinkurl{http://ascl.net/#1}}}
\providecommand{\doarXiv}[1]{\href{https://arxiv.org/abs/#1}{\nolinkurl{https://arxiv.org/abs/#1}}}

\bibitem[{J. {Birn} {et~al.}(2001){Birn}, {Drake}, {Shay}, {Rogers}, {Denton},
  {Hesse}, {Kuznetsova}, {Ma}, {Bhattacharjee}, {Otto}, \&
  {Pritchett}}]{birn+01}
{Birn}, J., {Drake}, J.~F., {Shay}, M.~A., {et~al.} 2001,
  \bibinfo{title}{{Geospace Environmental Modeling (GEM) magnetic reconnection
  challenge},} \jgr, 106, 3715, \dodoi{10.1029/1999JA900449}

\bibitem[{J.~L. {Burch} {et~al.}(2016){Burch}, {Torbert}, {Phan}, {Chen},
  {Moore}, {Ergun}, {Eastwood}, {Gershman}, {Cassak}, {Argall}, {Wang},
  {Hesse}, {Pollock}, {Giles}, {Nakamura}, {Mauk}, {Fuselier}, {Russell},
  {Strangeway}, {Drake}, {Shay}, {Khotyaintsev}, {Lindqvist}, {Marklund},
  {Wilder}, {Young}, {Torkar}, {Goldstein}, {Dorelli}, {Avanov}, {Oka},
  {Baker}, {Jaynes}, {Goodrich}, {Cohen}, {Turner}, {Fennell}, {Blake},
  {Clemmons}, {Goldman}, {Newman}, {Petrinec}, {Trattner}, {Lavraud}, {Reiff},
  {Baumjohann}, {Magnes}, {Steller}, {Lewis}, {Saito}, {Coffey}, \&
  {Chandler}}]{burch+16b}
{Burch}, J.~L., {Torbert}, R.~B., {Phan}, T.~D., {et~al.} 2016,
  \bibinfo{title}{{Electron-scale measurements of magnetic reconnection in
  space},} Science, 352, aaf2939, \dodoi{10.1126/science.aaf2939}

\bibitem[{P.~A. {Cassak}(2011){Cassak}}]{cassak11}
{Cassak}, P.~A. 2011, \bibinfo{title}{{Theory and simulations of the scaling of
  magnetic reconnection with symmetric shear flow},} Physics of Plasmas, 18,
  072106, \dodoi{10.1063/1.3602859}

\bibitem[{P.~A. {Cassak} \& A. {Otto}(2011){Cassak} \& {Otto}}]{cassak+11}
{Cassak}, P.~A., \& {Otto}, A. 2011, \bibinfo{title}{{Scaling of the magnetic
  reconnection rate with symmetric shear flow},} Physics of Plasmas, 18,
  074501, \dodoi{10.1063/1.3609771}

\bibitem[{S. {Chandrasekhar}(1961){Chandrasekhar}}]{chandrasekhar61}
{Chandrasekhar}, S. 1961, {Hydrodynamic and hydromagnetic stability}

\bibitem[{A. Chasapis {et~al.}(2018)Chasapis, Matthaeus, Parashar, Wan,
  Haggerty, Pollock, Giles, Paterson, Dorelli, Gershman, Torbert, Russell,
  Lindqvist, Khotyaintsev, Moore, Ergun, \& Burch}]{chasapis+18b}
Chasapis, A., Matthaeus, W.~H., Parashar, T.~N., {et~al.} 2018,
  \bibinfo{title}{In Situ Observation of Intermittent Dissipation at Kinetic
  Scales in the Earth's Magnetosheath,} Astrophysical Journal Letters, 856,
  L19, \dodoi{10.3847/2041-8213/aaadf8}

\bibitem[{C.~E. {Doss} {et~al.}(2016){Doss}, {Cassak}, \& {Swisdak}}]{doss+16}
{Doss}, C.~E., {Cassak}, P.~A., \& {Swisdak}, M. 2016,
  \bibinfo{title}{{Particle-in-cell simulation study of the scaling of
  asymmetric magnetic reconnection with in-plane flow shear},} Physics of
  Plasmas, 23, 082107, \dodoi{10.1063/1.4960324}

\bibitem[{J.~F. {Drake} \& M. {Swisdak}(2012){Drake} \& {Swisdak}}]{drake+12}
{Drake}, J.~F., \& {Swisdak}, M. 2012, \bibinfo{title}{{Ion Heating and
  Acceleration During Magnetic Reconnection Relevant to the Corona},} \ssr,
  172, 227, \dodoi{10.1007/s11214-012-9903-3}

\bibitem[{J.~W. Dungey(1961)Dungey}]{dungey61}
Dungey, J.~W. 1961, \bibinfo{title}{Interplanetary Magnetic Field and the
  Auroral Zones,} Phys. Rev. Lett., 6, 47, \dodoi{10.1103/PhysRevLett.6.47}

\bibitem[{R.~E. {Ergun} {et~al.}(2022){Ergun}, {Pathak}, {Usanova}, {Qi}, {Vo},
  {Burch}, {Schwartz}, {Torbert}, {Ahmadi}, {Wilder}, {Chasipis}, {Newman},
  {Stawarz}, {Hesse}, {Turner}, \& {Gershman}}]{ergun+22}
{Ergun}, R.~E., {Pathak}, N., {Usanova}, M.~E., {et~al.} 2022,
  \bibinfo{title}{{Observation of Magnetic Reconnection in a Region of Strong
  Turbulence},} \apjl, 935, L8, \dodoi{10.3847/2041-8213/ac81d4}

\bibitem[{S.~A. {Fuselier} {et~al.}(2005){Fuselier}, {Trattner}, {Petrinec},
  {Owen}, \& {R{\`e}me}}]{fuselier+05}
{Fuselier}, S.~A., {Trattner}, K.~J., {Petrinec}, S.~M., {Owen}, C.~J., \&
  {R{\`e}me}, H. 2005, \bibinfo{title}{{Computing the reconnection rate at the
  Earth's magnetopause using two spacecraft observations},} Journal of
  Geophysical Research (Space Physics), 110, A06212,
  \dodoi{10.1029/2004JA010805}

\bibitem[{C.~A. {Giai} {et~al.}(2024){Giai}, {Haggerty}, {Shay}, {Cassak}, \&
  {Davis}}]{giai+24}
{Giai}, C.~A., {Haggerty}, C.~C., {Shay}, M.~A., {Cassak}, P.~A., \& {Davis},
  Z.~K. 2024, \bibinfo{title}{{Suppression of Collisionless Magnetic
  Reconnection in the High Ion {\ensuremath{\beta}}, Strong Guide Field
  Limit},} \apj, 977, 218, \dodoi{10.3847/1538-4357/ad9274}

\bibitem[{J.~T. {Gosling}(2007){Gosling}}]{gosling+07b}
{Gosling}, J.~T. 2007, \bibinfo{title}{{Observations of Magnetic Reconnection
  in the Turbulent High-Speed Solar Wind},} \apjl, 671, L73,
  \dodoi{10.1086/524842}

\bibitem[{J.~T. {Gosling} {et~al.}(2007){Gosling}, {Eriksson}, {Phan},
  {Larson}, {Skoug}, \& {McComas}}]{gosling+07a}
{Gosling}, J.~T., {Eriksson}, S., {Phan}, T.~D., {et~al.} 2007,
  \bibinfo{title}{{Direct evidence for prolonged magnetic reconnection at a
  continuous x-line within the heliospheric current sheet},} \grl, 34, L06102,
  \dodoi{10.1029/2006GL029033}

\bibitem[{J.~T. {Gosling} {et~al.}(2005){Gosling}, {Skoug}, {McComas}, \&
  {Smith}}]{gosling+05}
{Gosling}, J.~T., {Skoug}, R.~M., {McComas}, D.~J., \& {Smith}, C.~W. 2005,
  \bibinfo{title}{{Magnetic disconnection from the Sun: Observations of a
  reconnection exhaust in the solar wind at the heliospheric current sheet},}
  \grl, 32, L05105, \dodoi{10.1029/2005GL022406}

\bibitem[{C.~C. Haggerty(2016)Haggerty}]{haggerty+16b}
Haggerty, C.~C. 2016, \bibinfo{title}{Ion and electron heating during magnetic
  reconnection in simulations,} PhD thesis, University of Delaware.
\newblock \url{http://adsabs.harvard.edu/abs/2016PhDT.......198H}

\bibitem[{C.~C. Haggerty {et~al.}(2017)Haggerty, Parashar, Matthaeus, Shay,
  Yang, Wan, Wu, \& Servidio}]{haggerty+17}
Haggerty, C.~C., Parashar, T.~N., Matthaeus, W.~H., {et~al.} 2017,
  \bibinfo{title}{Exploring the statistics of magnetic reconnection X-points in
  kinetic particle-in-cell turbulence,} Physics of Plasmas, 24, 102308,
  \dodoi{10.1063/1.5001722}

\bibitem[{C.~C. Haggerty {et~al.}(2018)Haggerty, Shay, Chasapis, Phan, Drake,
  Malakit, Cassak, \& Kieokaew}]{haggerty+18}
Haggerty, C.~C., Shay, M.~A., Chasapis, A., {et~al.} 2018, \bibinfo{title}{The
  reduction of magnetic reconnection outflow jets to sub-Alfvénic speeds,}
  Physics of Plasmas, 25, 102120, \dodoi{10.1063/1.5050530}

\bibitem[{C.~C. Haggerty {et~al.}(2015)Haggerty, Shay, Drake, Phan, \&
  McHugh}]{haggerty+15}
Haggerty, C.~C., Shay, M.~A., Drake, J.~F., Phan, T.~D., \& McHugh, C.~T. 2015,
  \bibinfo{title}{The competition of electron and ion heating during magnetic
  reconnection,} Geophysical Research Letters, 42, 9657,
  \dodoi{10.1002/2015GL065961}

\bibitem[{H. {Hasegawa} {et~al.}(2016){Hasegawa}, {Kitamura}, {Saito}, {Nagai},
  {Shinohara}, {Yokota}, {Pollock}, {Giles}, {Dorelli}, {Gershman}, {Avanov},
  {Kreisler}, {Paterson}, {Chandler}, {Coffey}, {Burch}, {Torbert}, {Moore},
  {Russell}, {Strangeway}, {Le}, {Oka}, {Phan}, {Lavraud}, {Zenitani}, \&
  {Hesse}}]{hasegawa+16}
{Hasegawa}, H., {Kitamura}, N., {Saito}, Y., {et~al.} 2016,
  \bibinfo{title}{{Decay of mesoscale flux transfer events during
  quasi-continuous spatially extended reconnection at the magnetopause},} \grl,
  43, 4755, \dodoi{10.1002/2016GL069225}

\bibitem[{A.~L. {La Belle-Hamer} {et~al.}(1995){La Belle-Hamer}, {Otto}, \&
  {Lee}}]{labelle-hamer+95}
{La Belle-Hamer}, A.~L., {Otto}, A., \& {Lee}, L.~C. 1995,
  \bibinfo{title}{{Magnetic reconnection in the presence of sheared flow and
  density asymmetry: Applications to the Earth's magnetopause},} \jgr, 100,
  11875, \dodoi{10.1029/94JA00969}

\bibitem[{X. {Li} \& Y.-H. {Liu}(2021){Li} \& {Liu}}]{li+21}
{Li}, X., \& {Liu}, Y.-H. 2021, \bibinfo{title}{{The Effect of Thermal Pressure
  on Collisionless Magnetic Reconnection Rate},} \apj, 912, 152,
  \dodoi{10.3847/1538-4357/abf48c}

\bibitem[{H. {Liang} {et~al.}(2024){Liang}, {Chen}, {Bessho}, \&
  {Ng}}]{liang+24}
{Liang}, H., {Chen}, L.-J., {Bessho}, N., \& {Ng}, J. 2024,
  \bibinfo{title}{{Impact of the Out-Of-Plane Flow Shear on Magnetic
  Reconnection at the Flanks of Earth's Magnetopause},} Journal of Geophysical
  Research (Space Physics), 129, e2024JA033154, \dodoi{10.1029/2024JA033154}

\bibitem[{T.~Z. Liu {et~al.}(2017)Liu, Angelopoulos, Hietala, \&
  Wilson}]{Liu+17}
Liu, T.~Z., Angelopoulos, V., Hietala, H., \& Wilson, III, L.~B. 2017,
  \bibinfo{title}{Statistical study of particle acceleration in the core of
  foreshock transients,} Journal of Geophysical Research (Space Physics), 122,
  7197, \dodoi{10.1002/2017JA024043}

\bibitem[{W.~H. {Matthaeus} \& S.~L. {Lamkin}(1986){Matthaeus} \&
  {Lamkin}}]{matthaeus+86}
{Matthaeus}, W.~H., \& {Lamkin}, S.~L. 1986, \bibinfo{title}{{Turbulent
  magnetic reconnection},} Physics of Fluids, 29, 2513,
  \dodoi{10.1063/1.866004}

\bibitem[{W.~H. {Matthaeus} {et~al.}(2020){Matthaeus}, {Yang}, {Wan},
  {Parashar}, {Bandyopadhyay}, {Chasapis}, {Pezzi}, \&
  {Valentini}}]{matthaeus+20}
{Matthaeus}, W.~H., {Yang}, Y., {Wan}, M., {et~al.} 2020,
  \bibinfo{title}{{Pathways to Dissipation in Weakly Collisional Plasmas},}
  \apj, 891, 101, \dodoi{10.3847/1538-4357/ab6d6a}

\bibitem[{R. {Mbarek} {et~al.}(2022){Mbarek}, {Haggerty}, {Sironi}, {Shay}, \&
  {Caprioli}}]{mbarek+22}
{Mbarek}, R., {Haggerty}, C., {Sironi}, L., {Shay}, M., \& {Caprioli}, D. 2022,
  \bibinfo{title}{{Relativistic Asymmetric Magnetic Reconnection},} \prl, 128,
  145101, \dodoi{10.1103/PhysRevLett.128.145101}

\bibitem[{M. {{\O}ieroset} {et~al.}(2024){{\O}ieroset}, {Phan}, {Drake},
  {Starkey}, {Fuselier}, {Cohen}, {Haggerty}, {Shay}, {Oka}, {Gershman},
  {Maheshwari}, {Burch}, {Torbert}, \& {Strangeway}}]{oieroset+24}
{{\O}ieroset}, M., {Phan}, T.~D., {Drake}, J.~F., {et~al.} 2024,
  \bibinfo{title}{{Scaling of Ion Bulk Heating in Magnetic Reconnection
  Outflows for the High-Alfv{\'e}n-speed and Low-{\ensuremath{\beta}} Regime in
  Earth's Magnetotail},} \apj, 971, 144, \dodoi{10.3847/1538-4357/ad6151}

\bibitem[{G. {Paschmann} {et~al.}(2013){Paschmann}, {{\O}ieroset}, \&
  {Phan}}]{paschmann+13}
{Paschmann}, G., {{\O}ieroset}, M., \& {Phan}, T. 2013,
  \bibinfo{title}{{In-Situ Observations of Reconnection in Space},} \ssr, 178,
  385, \dodoi{10.1007/s11214-012-9957-2}

\bibitem[{T.~D. {Phan} {et~al.}(1996){Phan}, {Paschmann}, \&
  {Sonnerup}}]{phan+96}
{Phan}, T.~D., {Paschmann}, G., \& {Sonnerup}, B.~U.~{\"O}. 1996,
  \bibinfo{title}{{Low-latitude dayside magnetopause and boundary layer for
  high magnetic shear 2. Occurrence of magnetic reconnection},} \jgr, 101,
  7817, \dodoi{10.1029/95JA03751}

\bibitem[{T.~D. {Phan} {et~al.}(2013){Phan}, {Shay}, {Gosling}, {Fujimoto},
  {Drake}, {Paschmann}, {Oieroset}, {Eastwood}, \& {Angelopoulos}}]{phan+13}
{Phan}, T.~D., {Shay}, M.~A., {Gosling}, J.~T., {et~al.} 2013,
  \bibinfo{title}{{Electron bulk heating in magnetic reconnection at Earth's
  magnetopause: Dependence on the inflow Alfv{\'e}n speed and magnetic shear},}
  \grl, 40, 4475, \dodoi{10.1002/grl.50917}

\bibitem[{T.~D. {Phan} {et~al.}(2014){Phan}, {Drake}, {Shay}, {Gosling},
  {Paschmann}, {Eastwood}, {Oieroset}, {Fujimoto}, \& {Angelopoulos}}]{phan+14}
{Phan}, T.~D., {Drake}, J.~F., {Shay}, M.~A., {et~al.} 2014,
  \bibinfo{title}{{Ion bulk heating in magnetic reconnection exhausts at
  Earth's magnetopause: Dependence on the inflow Alfv{\'e}n speed and magnetic
  shear angle},} \grl, 41, 7002, \dodoi{10.1002/2014GL061547}

\bibitem[{T.~D. Phan {et~al.}(2018)Phan, Eastwood, Shay, Drake, Sonnerup,
  Fujimoto, Cassak, {\O}ieroset, Burch, Torbert, Rager, Dorelli, Gershman,
  Pollock, Pyakurel, Haggerty, Khotyaintsev, Lavraud, Saito, Oka, Ergun,
  Retino, Le~Contel, Argall, Giles, Moore, Wilder, Strangeway, Russell,
  Lindqvist, \& Magnes}]{phan+18}
Phan, T.~D., Eastwood, J.~P., Shay, M.~A., {et~al.} 2018,
  \bibinfo{title}{Electron magnetic reconnection without ion coupling in
  Earth's turbulent magnetosheath,} Nature, 557, 202,
  \dodoi{10.1038/s41586-018-0091-5}

\bibitem[{T.~D. {Phan} {et~al.}(2020){Phan}, {Bale}, {Eastwood}, {Lavraud},
  {Drake}, {Oieroset}, {Shay}, {Pulupa}, {Stevens}, {MacDowall}, {Case},
  {Larson}, {Kasper}, {Whittlesey}, {Szabo}, {Korreck}, {Bonnell}, {de Wit},
  {Goetz}, {Harvey}, {Horbury}, {Livi}, {Malaspina}, {Paulson}, {Raouafi}, \&
  {Velli}}]{phan+20}
{Phan}, T.~D., {Bale}, S.~D., {Eastwood}, J.~P., {et~al.} 2020,
  \bibinfo{title}{{Parker Solar Probe In Situ Observations of Magnetic
  Reconnection Exhausts during Encounter 1},} \apjs, 246, 34,
  \dodoi{10.3847/1538-4365/ab55ee}

\bibitem[{A. {Retin{\`o}} {et~al.}(2007){Retin{\`o}}, {Sundkvist}, {Vaivads},
  {Mozer}, {Andr{\'e}}, \& {Owen}}]{retino+07}
{Retin{\`o}}, A., {Sundkvist}, D., {Vaivads}, A., {et~al.} 2007,
  \bibinfo{title}{{In situ evidence of magnetic reconnection in turbulent
  plasma},} Nature Physics, 3, 236, \dodoi{10.1038/nphys574}

\bibitem[{S. {Servidio} {et~al.}(2009){Servidio}, {Matthaeus}, {Shay},
  {Cassak}, \& {Dmitruk}}]{servidio+09}
{Servidio}, S., {Matthaeus}, W.~H., {Shay}, M.~A., {Cassak}, P.~A., \&
  {Dmitruk}, P. 2009, \bibinfo{title}{{Magnetic Reconnection in Two-Dimensional
  Magnetohydrodynamic Turbulence},} \prl, 102, 115003,
  \dodoi{10.1103/PhysRevLett.102.115003}

\bibitem[{M.~A. {Shay} {et~al.}(2001){Shay}, {Drake}, {Rogers}, \&
  {Denton}}]{shay+01}
{Shay}, M.~A., {Drake}, J.~F., {Rogers}, B.~N., \& {Denton}, R.~E. 2001,
  \bibinfo{title}{{Alfv{\'e}nic collisionless magnetic reconnection and the
  Hall term},} \jgr, 106, 3759, \dodoi{10.1029/1999JA001007}

\bibitem[{M.~A. Shay {et~al.}(2018)Shay, Haggerty, Matthaeus, Parashar, Wan, \&
  Wu}]{shay+18}
Shay, M.~A., Haggerty, C.~C., Matthaeus, W.~H., {et~al.} 2018,
  \bibinfo{title}{Turbulent heating due to magnetic reconnection,} Physics of
  Plasmas, 25, 012304, \dodoi{10.1063/1.4993423}

\bibitem[{M.~A. {Shay} {et~al.}(2014){Shay}, {Haggerty}, {Phan}, {Drake},
  {Cassak}, {Wu}, {Oieroset}, {Swisdak}, \& {Malakit}}]{shay+14}
{Shay}, M.~A., {Haggerty}, C.~C., {Phan}, T.~D., {et~al.} 2014,
  \bibinfo{title}{{Electron heating during magnetic reconnection: A simulation
  scaling study},} Physics of Plasmas, 21, 122902, \dodoi{10.1063/1.4904203}

\bibitem[{T.~W. {Speiser}(1965){Speiser}}]{speiser65}
{Speiser}, T.~W. 1965, \bibinfo{title}{{Particle Trajectories in Model Current
  Sheets, 1, Analytical Solutions},} \jgr, 70, 4219,
  \dodoi{10.1029/JZ070i017p04219}

\bibitem[{J.~E. Stawarz {et~al.}(2022)Stawarz, Eastwood, Phan, Gingell,
  Pyakurel, Shay, Robertson, Russell, \& Le~Contel}]{stawarz+22}
Stawarz, J.~E., Eastwood, J.~P., Phan, T.~D., {et~al.} 2022,
  \bibinfo{title}{Turbulence-driven magnetic reconnection and the magnetic
  correlation length: Observations from Magnetospheric Multiscale in Earth's
  magnetosheath,} Physics of Plasmas, 29, 012302, \dodoi{10.1063/5.0071106}

\bibitem[{M. {Wan} {et~al.}(2013){Wan}, {Matthaeus}, {Servidio}, \&
  {Oughton}}]{wan+13}
{Wan}, M., {Matthaeus}, W.~H., {Servidio}, S., \& {Oughton}, S. 2013,
  \bibinfo{title}{{Generation of X-points and secondary islands in 2D
  magnetohydrodynamic turbulence},} Physics of Plasmas, 20, 042307,
  \dodoi{10.1063/1.4802985}

\bibitem[{J. {Wang} {et~al.}(2008){Wang}, {Wang}, \& {Xiao}}]{wang+08}
{Wang}, J., {Wang}, X., \& {Xiao}, C. 2008, \bibinfo{title}{{Out-of-plane
  bipolar and quadrupolar magnetic fields generated by shear flows in
  two-dimensional resistive reconnection},} Physics Letters A, 372, 4614,
  \dodoi{10.1016/j.physleta.2008.04.048}

\bibitem[{J. {Wang} {et~al.}(2012){Wang}, {Xiao}, \& {Wang}}]{wang+12}
{Wang}, J., {Xiao}, C., \& {Wang}, X. 2012, \bibinfo{title}{{Effects of
  out-of-plane shear flows on fast reconnection in a two-dimensional Hall
  magnetohydrodynamics model},} Physics of Plasmas, 19, 032905,
  \dodoi{10.1063/1.3697561}

\bibitem[{L. {Wang} {et~al.}(2015){Wang}, {Wang}, {Wang}, \& {Liu}}]{wang+15}
{Wang}, L., {Wang}, X.-G., {Wang}, X.-Q., \& {Liu}, Y. 2015,
  \bibinfo{title}{{Asymmetric magnetic reconnection with out-of-plane shear
  flows in a two dimensional hybrid model},} Physics of Plasmas, 22, 052110,
  \dodoi{10.1063/1.4919965}

\bibitem[{J. {Yoo} {et~al.}(2014){Yoo}, {Yamada}, {Ji}, {Jara-Almonte}, \&
  {Myers}}]{yoo+14}
{Yoo}, J., {Yamada}, M., {Ji}, H., {Jara-Almonte}, J., \& {Myers}, C.~E. 2014,
  \bibinfo{title}{{Bulk ion acceleration and particle heating during magnetic
  reconnection in a laboratory plasmaa)},} Physics of Plasmas, 21, 055706,
  \dodoi{10.1063/1.4874331}

\bibitem[{A. {Zeiler} {et~al.}(2002){Zeiler}, {Biskamp}, {Drake}, {Rogers},
  {Shay}, \& {Scholer}}]{zeiler+02}
{Zeiler}, A., {Biskamp}, D., {Drake}, J.~F., {et~al.} 2002,
  \bibinfo{title}{{Three-dimensional particle simulations of collisionless
  magnetic reconnection},} Journal of Geophysical Research (Space Physics),
  107, 1230, \dodoi{10.1029/2001JA000287}

\bibitem[{M. {Zhou} {et~al.}(2021){Zhou}, {Man}, {Deng}, {Pang},
  {Khotyaintsev}, {Lapenta}, {Yi}, {Zhong}, \& {Ma}}]{zhou+21}
{Zhou}, M., {Man}, H.~Y., {Deng}, X.~H., {et~al.} 2021,
  \bibinfo{title}{{Observations of Secondary Magnetic Reconnection in the
  Turbulent Reconnection Outflow},} \grl, 48, e91215,
  \dodoi{10.1029/2020GL091215}

\end{thebibliography}

\end{document}